\documentclass[prd,twocolumn,tightenlines,showpacs,nofootinbib,amsfonts,amssymb,amsmath]{revtex4}
\usepackage{graphicx}
\begin{document}
\newcommand{\etal}{{\it et al.}}
\newcommand{\bx}{{\bf x}}
\newcommand{\bn}{{\bf n}}
\newcommand{\bk}{{\bf k}}
\newcommand{\dd}{{\rm d}}
\newcommand{\dslash}{D\!\!\!\!/}
\def\ga{\mathrel{\raise.3ex\hbox{$>$\kern-.75em\lower1ex\hbox{$\sim$}}}}
\def\la{\mathrel{\raise.3ex\hbox{$<$\kern-.75em\lower1ex\hbox{$\sim$}}}}
\def\beq{\begin{equation}}
\def\eeq{\end{equation}}

\leftline{UMN--TH--3041/12}
\leftline{FTPI--MINN--12/14}

\vskip-2cm
\title{Where are the Walls?}

\author{Keith A. Olive$^{1,2}$, Marco Peloso$^2$
and Adam J. Peterson$^{2}$}

\affiliation{
${^1}$ William I. Fine Theoretical Physics Institute, 
University of Minnesota, Minneapolis, 55455, (USA) \\
${^2}$ School of Physics and Astronomy,
University of Minnesota, Minneapolis, 55455 (USA)\\
}
\vspace*{2cm}
\begin{abstract}
The reported spatial variation in the fine-structure constant at high redshift, if physical,  could 
be due to the presence of dilatonic domains, and one or more domain walls inside our horizon.
An  absorption spectrum of an object in a different domain from our own would be characterized by a different value of $\alpha$.  We show that while a single wall solution is statically comparable to a dipole fit, and is a big improvement over a weighted mean (despite adding 3 parameters), a two-wall solution is a far better fit (despite adding 3 parameters over the single wall solution). We derive a simple model accounting for the two-domain wall solution. The goodness of these fits is however dependent on 
the extra random error which was argued to account for the large scatter in most of the data. When this error is omitted, all the above solutions are poor fits to the data. When included, the solutions that exhibit a spatial dependence agree with the data much more significantly than the Standard Model; however, the Standard Model itself is not a terrible fit to the data, having a p-value of $ \sim 20 \%$.
\end{abstract}
 \date{April 2012}
 \maketitle

\section{Introduction}

The universality of fundamental constants is one of the underlying
tenets of physics. Of course some of these constants, like the 
fine-structure constant, $\alpha$, may be dynamical and their values
may be the result of an expectation value of some scalar field.
Thus testing this universality is a bridge to physics beyond the Standard Model~\cite{jp-revue}.

Claims of a temporal variation in $\alpha$  from observations of quasar absorption spectra
have been extended to include a possible spatial variation as well \cite{newwebb}.
The initial indications of a temporal variation in $\alpha$\ made use of
the many-multiplet method \cite{Webb} and sparked an enormous amount of 
theoretical activity in attempts to explain it~\cite{massgrave,SBM,seealso,op1,cststring}. 
The Keck/Hires data which yielded a statistically
significant trend with
$\Delta \alpha / \alpha = (-0.54 \pm 0.12) \times 10^{-5}$ over a redshift 
range $0.5 \la z \la 3.0$ (the minus sign indicates a smaller value of 
$\alpha$ in the past).  
Subsequent studies based on VLT data using the same method
have shown $\Delta \alpha$ to be consistent with zero \cite{Petitjean,quast}.
Of course these results can be made compatible if there is a spatial variation in 
$\alpha$.

Variations of $\alpha$ observed in individual absorbers using the many-multiplet method
or more generally temporal variations, could be due to systematics related to the
astrophysical assumptions made regarding each absorber.  For example,
it is generally assumed that the abundance ratio of Mg isotopes take
their terrestrial values. However, even a slight enhancement in $^{25,26}$Mg/$^{24}$Mg
could nullify many of  the observed variations. Furthermore, such an enhancement in the 
heavier Mg isotopes could be explained by an earlier population of intermediate mass
stars \cite{sys}.

Results from a recent VLT survey of 153 absorbers was performed \cite{newwebb}, and
taken together with the Keck data, leads to a large sample of 293 absorbers
across the sky. However, it was shown, that there is a statistically significant spatial
variation as demonstrated by a dipole fit to the data 
\beq
{ \Delta \alpha \over \alpha} = m + A \, \cos \left( \vartheta \right) \, ,
\label{dipole}
\eeq
which can not be accounted for by the systematic uncertainty due to the Mg isotopic abundances.
In Eq. (\ref{dipole}), 
$m$ is the monopole term, $A$ is the magnitude of the dipole,
and $\vartheta$ is the angle between the directions of the absorbers and the dipole. 
The template therefore has four parameters, $m$, $A$, and the two angles
which specify the direction of the dipole. There is a marked improvement
to the fit (using the full data set) using the dipole rather than a simple 
weighted mean (monopole) despite adding three parameters.

If the fine-structure constant does depend on some
dynamical scalar field, $\phi$, it may well be space-time dependent.
On cosmological scales, it is usually thought
that the time variation dominates
over spatial fluctuations, as suggested by most models.  
For example, suppose we couple a scalar field to electromagnetism 
through $\frac{B_{F}(\phi)}{4}F_{\mu\nu}F^{\mu\nu}$,
$F_{\mu\nu}$ being the Faraday tensor and $B_F$ an arbitrary function
of $\phi$. This will necessarily induce a coupling to matter which is generated radiatively 
if not present at the tree level (see below).
The equation of motion for the scalar field simply takes the form
\beq
\Box \phi +
\frac{\partial V_{\rm eff}}{\partial \phi}  =0,
\label{fieldeq} 
\eeq
where $V_{\rm eff}$ includes the self interactions of $\phi$ as well as any couplings to matter.
If the Lagrangian contains a term $B_{N}(\phi) m_N {\bar N} N$,
then the coupling to matter is effectively density dependent, and
could serve as the source of spatial variations through
\beq
\Box \phi + m_\phi^2 \phi =  B_N'(\phi) \rho_N ,
\eeq
where $m_\phi$ is the scalar mass and $\rho_N$ is the baryon energy density, as is the
case for the chameleon mechanism~\cite{chameleon}.
However, the density dependent shifts from the homogeneous solution
are typically extremely small except perhaps in the vicinity of 
a neutron star \cite{ellisolive,op2,barrow}. In contrast, temporal variations 
are relatively easy to achieve particularly over cosmological time scales,
as long as the field remains light.

Thus, we know of no physical or field theoretic model which could
produce a dipole accounting for the observed spatial variations.
Instead, we proposed \cite{opu} to invoke the existence of a spatial discontinuity of the
fine structure constant due to the existence of a domain
wall crossing our Hubble volume (this idea was further studied in \cite{wall-alpha}). 
In this case, $\alpha$ would 
take two values, $\alpha_+$ (the larger value of $\alpha$) 
on our side of the wall and the second
$\alpha_-$ at high redshifts on the other side of the wall.  This implies
that local constraints~\cite{jp-revue} on the variation of $\alpha$ such as atomic clocks,
Oklo and meteoritic dating will be trivially satisfied.

The simplest way to implement this idea is to consider
the following theory
\begin{eqnarray}
 S =\int \left[ \frac{1}{2}M_p^2 R -\frac{1}{2}(\partial_\mu\phi)^2
 -V(\phi)-\frac{1}{4}B_F(\phi) F_{\mu\nu}^2 \right. \nonumber \\ 
\left. - \sum_j  i\bar\psi_j\dslash\psi_j - B_j(\phi) m_j \bar\psi_j\psi_j
 \right]\sqrt{-g}\dd^4 x,
\label{lagrangian}
\end{eqnarray}
where $M_p^{-2}=8\pi G$ is the reduced Planck mass. The scalar field $\phi$ is assumed to have a simple 
quartic potential
\begin{equation}
 V(\phi) = \frac{1}{4}\lambda (\phi^2 - \eta^2)^2 \, ,
 \label{quartic}
\end{equation}
and a coupling to the Faraday tensor  
as well as to the fermions $\psi_j$.
The coupling functions $B_i$ are assumed
to be of the form
\begin{equation}
 B_i(\phi) = \exp\left({\xi_i \frac{\phi}{M_*}}\right)\simeq 1 + \xi_i \frac{\phi}{M_*},
\label{B-exp}
\end{equation}
where the coefficients $\xi_i$ are constant and $M_*$ is a mass scale. This model
depends on the parameters $(\lambda,M_*,\eta,\xi_F,\xi_i)$ and we shall assume here that,
at tree-level, only $\xi_F$ is non-vanishing. 
Nevertheless, the scalar field inevitably  couples to nucleons radiatively through 
$\xi_N =  m_N^{-1} \langle N|(\xi_F/4) F_{\mu\nu}^2| N \rangle$~\cite{op1}. 
This yields $\xi_{\rm p} = -0.0007 \xi_F$ and $\xi_{\rm n} = 0.00015 \xi_F$~\cite{gl} for the
proton and neutron,  respectively.
Since most  baryons in the universe are protons, 
we shall take $\xi_N = \xi_{\rm p}$ for simplicity in our estimates. 

The main difference between the model studied here and previous models is that the scalar field
is assumed to be heavy so that it is stabilized, hence we do not
expect any local violation of the equivalence principle. Indeed, the current model
does not exhibit any temporal variation of constants once the phase transition has occurred.
The resulting shift in the value of $\alpha$ on the other side of the wall is easily determined \cite{opu}
\begin{equation}
 \frac{\Delta \alpha}{\alpha} \simeq  2\xi_F  \,  \frac{\eta}{M_*} .
\end{equation}
For simplicity, we shall assume  $\eta = M_*$, so 
that $\xi_F \simeq 10^{-6}$ is required to match the data. 
Then for $\lambda \sim 1$ and $\eta \sim 1$ MeV, one can show that such 
a wall makes only a tiny contribution to the overall energy density
( $\Omega_{\rm wall} \sim \left(\frac{\eta}{100\,{\rm MeV}}\right)^{3}$) \cite{obswall},
it is cosmologically stable, and is compatible with microwave background \cite{CMBstring} and other astrophysical
constraints. Furthermore, due to the dynamics of the phase transition
producing the wall, we expect to be left with order one large wall per Hubble radius \cite{walldyn}
which is moving slowly towards us \cite{opu}.

Given the large data set available \cite{newwebb,Webb}, it is possible to be
more quantitative concerning the wall. We can in fact use the data to determine
the position of the wall, and the potential drop across it. We can further make 
a statistical judgement as to whether the wall is an improvement over
a simple temporal variation (a monopole fit), and compare it directly to the dipole
fit found in \cite{newwebb}.  As we will see below, while in ideal circumstances,
the wall is an approximation to a dipole, the direction of the wall only lines
up well with the dipole when using the Keck data alone. For the full data set,
best fit position of the wall is not aligned with the dipole.  Nevertheless, 
the fit for the wall as determined by a $\chi^2$ analysis is comparable to that of the dipole.
While less significant than the Keck data, 
the VLT data shows a tendency for positive variations of $\alpha$.
We show that a two-wall solution (despite requiring three additional parameters)
is a far superior fit to the data than either the dipole or single wall solution.

In the next Section,  we present a model with three dilatonic solutions, which is the basis for the two-wall fits of the data discussed below. This model is a simple extension of the model of \cite{opu}. In Section \ref{data}, we briefly describe the data that we use. In Section \ref{fit-wall},  we outline the 
key algorithm used in the walls fit, used to determine in which vacuum an absorber lies. In Section 
\ref{fit-results}, we present the results of our fits. The significance for a spatial dependence in the data, and the orientation of the dipole and wall fits are  discussed in Section \ref{fit-significance}.  Our conclusions are summarized in section \ref{summary}.

\section{The two-wall model} \label{twowalls}

In many extensions of the Standard Model, the value of $\alpha$ is a function of the dilaton field, 
as indicated in (\ref{lagrangian}). In general, $\phi$ is the real part of a complex field $\Phi$. In \cite{opu}, it was suggested that the potential of $\phi$ has two minima, and that we live in one of them, while some of the observed absorbers live in the other vacuum. In this paper, we  also fit the data against a two-wall model. In this fit, absorbers have either the same value  $\alpha_0$ that we measured on Earth, or the value $\alpha_0 \left( 1 +  \Delta \right)$, or $\alpha_0 \left( 1 - \Delta \right)$. In this Section, we briefly indicate how the model of  \cite{opu} can be modified to have three dilatonic vacua. 

The relevant term in the action are 
\begin{eqnarray}
{\cal L} & = &- \vert \partial_\mu \Phi \vert^2 - V \left( \Phi \right) - \frac{1}{4} B_F \left( \phi \right) F_{\mu\nu}^2
\;\;\;,\;\;\; \phi = {\rm Re } \, \Phi \nonumber\\
V & = & \lambda \left( \vert \Phi \vert^2 - \frac{\eta^2}{2} \right)^2 - \sqrt{2} \, i \epsilon \left( \Phi^3 - \Phi^{* 3} \right) + V_0 \, ,
\label{lag-3vac}
\end{eqnarray}
where the three parameters $\lambda ,\, \eta ,\, \epsilon$ are positive and real numbers. We note that $\lambda$ is dimensionless, while $\eta$ and $\epsilon$ have mass dimension one. The constant $V_0$ is chosen such that the potential vanishes at its minimum. For computational simplicity, we assume that $\epsilon \ll \lambda \, \eta$, so that the second term in the potential can be treated as a perturbation. The potential has the three minima
\begin{eqnarray}
\Phi_n & = & \frac{\eta}{\sqrt{2}} \, R_0 \, {\rm e}^{i \left(  \frac{2 \pi}{3}  n - \frac{\pi}{6} 
\right)} \;\;,\;\; n = 1 ,\, 2 ,\, 3 \nonumber\\
R_0 & = & \sqrt{1 + \frac{9 \, \epsilon^2}{4 \, \eta^2 \, \lambda^2}} +  \frac{3 \, \epsilon}{2 \, \eta \, \lambda} = 1 + {\rm O } \left( \frac{ \epsilon }{ \lambda \, \eta } \right) \, .
\label{3min}
\end{eqnarray} 

\begin{figure}
\centerline{
\includegraphics[width=0.5\textwidth]{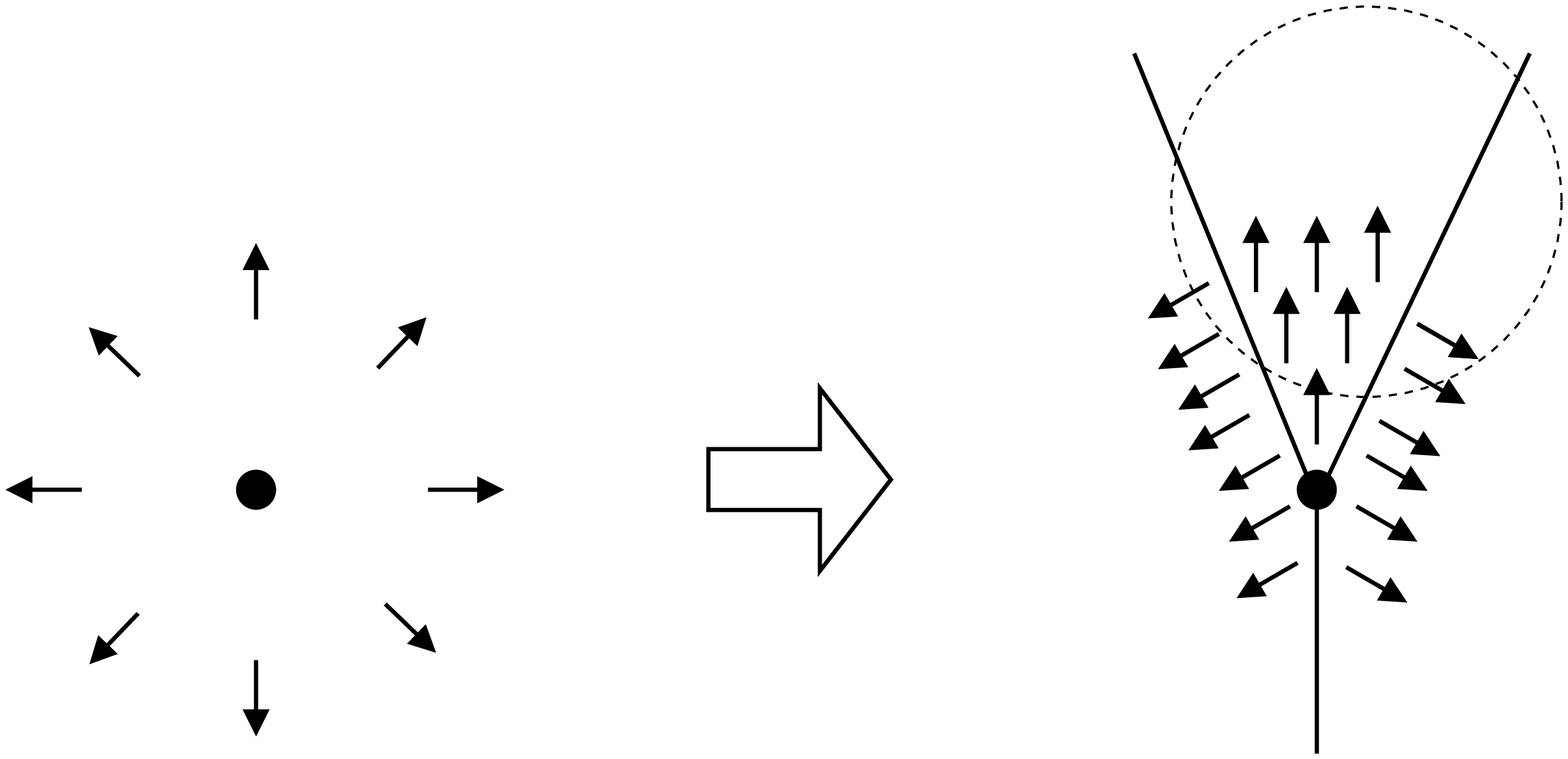}
}
\caption{The left side of the Figure illustrates a cosmic  string formed at the first stage of symmetry breaking; the arrows indicate the complex value taken by $\Phi$ in different spatial positions. The central dot represents the string (which extends out of the page). At the second stage of symmetry breaking, $\Phi$ settles in the different minima (\ref{3min}) inside  separated domains as seen in the right side of the Figure. The solid lines represent  sections of the walls separating the domains. This part of the Figure also illustrates a typical wall configuration in our fits:  our location is at the center of the dotted circle, which represents the typical radial distance to the Keck and VLT absorbers.}
\label{fig:string-wall}
\end{figure}

The phases of the three minima correspond to the complex vectors shown in the right half of Figure \ref{fig:string-wall}. We assume that  we live in the $n=1$ vacuum. To evaluate the values of $\alpha$ in the other vacua, we assume that $B$ is still given by  eq. (\ref{B-exp}), with $\phi = {\rm Re } \, \Phi$, and expand it for small $\xi$. We obtain
\begin{eqnarray}
&&\frac{\Delta \alpha_1}{\alpha} = 0 \;\;,\;\;
\frac{\Delta \alpha_2}{\alpha} = \Delta \;\;,\;\;
\frac{\Delta \alpha_3}{\alpha} = - \,  \Delta \nonumber\\
&&\Delta = \xi_N \,  \frac{\sqrt{3}}{2 \, \sqrt{2}} \, \frac{ \eta }{ M_* }
\left[  1 + {\rm O } \left( \frac{ \epsilon }{ \lambda \, \eta } \right) \right] \, .
\end{eqnarray}

For $\epsilon \ll  \lambda \eta$,  the radial excitations of $\Phi$ around the minima are much more massive than the angular excitations. At energies below the mass of the massive radial excitations, we have a consistent physical description which includes only the light degrees of freedom. At leading order in $\epsilon$, the canonically normalized field, $\theta_c$,  corresponding to the light excitations is 
\begin{equation}
\Phi \equiv \frac{\eta}{\sqrt{2}} \, {\rm e}^{i \frac{\theta_c}{\eta}} + {\rm O } \left( \epsilon\right) \, ,
\end{equation} 
and is controlled by the lagrangian
\begin{equation}
{\cal L} = -  \frac{1}{2} \left( \partial \theta_c \right)^2 - \epsilon \,  \eta^3 \left[ \sin  \left( \frac{3 \, \theta_c}{\eta} \right) + 1 \right] + {\rm O } \left( \epsilon^2 \right) \, .
\label{lag-dom}
\end{equation}

In the following, we assume $\lambda = {\rm O } \left( 1 \right)$ for definiteness. 
At high temperatures, the potential (\ref{lag-3vac}) is modified by thermal effects. At temperatures greater than ${\rm O } \left( \eta \right)$ the model is in the unbroken phase, namely the potential is minimized at $\Phi = 0$. Below this temperature, the nearly U(1) invariant set of minima $\vert \Phi \vert \simeq \frac{\eta}{\sqrt{2}}$ appears. This residual U(1) symmetry is preserved as long as thermal fluctuations are energetic enough to cross the potential barrier between the three vacua (\ref{3min}); this is the case until  $T \simeq {\rm O } \left(  \left( \frac{\epsilon}{\eta} \right)^{1/4} \, \eta \right) $. Below this temperature, the system is in the fully broken phase.

Therefore, for $\epsilon \ll \eta$, symmetry breaking occurs in two stages. Cosmic strings are formed at the first stage. Outside each string, $\Phi$ has fixed magnitude and variable phase. As the temperature decreases, domains of the three vacua 
(\ref{3min}) form; the different domains are separated by a domain configuration with $\vert \Phi \vert = \frac{\eta}{\sqrt{2}}  + {\rm O } \left( \epsilon \right)$, and with a phase interpolating between the values in the two domains. As illustrated in Figure \ref{fig:string-wall}, at least three domain walls stream off each string (more domains will be present for strings characterized by a higher winding number).

For  a planar and static wall perpendicular to the $z$ axis, it is convenient to define
\begin{equation}
{\tilde \theta} \equiv \frac{3 \theta_c}{\eta} - \frac{5 \pi}{2}  \;\;\;,\;\;\; 
{\tilde z} \equiv 3 \sqrt{\epsilon \, \eta} \, z  \, ,
\label{rescale}
\end{equation}
so that the action is manifestly put in the form of the  sine-Gordon model. Indeed, starting from 
(\ref{lag-dom}), and performing these redefinitions, we obtain the domain wall action
\begin{eqnarray}
S & = & \int d t \, d x \, d y \frac{1}{3} \, \sqrt{\frac{\epsilon}{\eta}} \, \eta^3 \int d {\tilde z} \left[ - \frac{1}{2} \left( \frac{d {\tilde \theta}}{d {\tilde z}} \right)^2 - {\tilde V } \right] \nonumber\\
{\tilde V} & \equiv & \cos \, {\tilde \theta} + 1 \, .
\end{eqnarray}
In these coordinates,  the three minima (\ref{3min}) correspond to the ${\tilde \theta} = - \pi ,\, \pi ,\, 3 \pi$
minima  of  ${\tilde V } $, respectively. The wall solution is obtained from a standard BPS procedure. 
The equation of motion $\frac{d^2 {\tilde \theta}}{d {\tilde z}^2} - \frac{d {\tilde  V}}{d {\tilde \theta}} = 0$ is integrated to give
\begin{equation}
\frac{1}{2} \left( \frac{d {\tilde \theta}}{d {\tilde z}} \right)^2 - {\tilde  V } \left( {\tilde  \theta } \right) = 0 \, .
\label{constrain}
\end{equation}
This equation can be further integrated to give (returning to the original coordinates)
\begin{equation}
\theta_c = \eta \left[ \frac{\pi}{2} + \frac{4}{3} \, \tan^{-1} \left( {\rm e}^{2 \sqrt{\epsilon \, \eta} \,  z} \right) \right] \, .
\end{equation}
This is the domain wall solution interpolating between the two minima $n=1$ (at $z=-\infty$) and $n=2$ (at $z=+\infty$) of (\ref{3min}). We note that the thickness of the wall is  of ${\rm O } \left( \frac{1}{\sqrt{\epsilon \, \eta}} \right)$.

The tension of the wall is most easily computed using the rescaling (\ref{rescale}), and using (\ref{constrain}), we obtain
\begin{eqnarray}
T  &=&  \frac{\rm energy}{\rm area}  =  \int dz \left[ \frac{1}{2} \left( \frac{d \theta}{d z} \right)^2 + V \right] \nonumber\\
&=& \frac{1}{3} \, \sqrt{\frac{\epsilon}{\eta}} \, \eta^3 \int_{-\pi}^\pi d {\tilde \theta} \, \sqrt{2 {\tilde V}} = 
\frac{8}{3} \sqrt{\frac{\epsilon}{\eta}} \, \eta^3 \, .
\label{tension}
\end{eqnarray}

The cosmological and astrophysical limits for the single wall model were discussed in Ref. \cite{opu}. This discussion can be readily extended to the current case. For example, the requirement that the CMB is not distorted by the walls translates into an upper limit \cite{opu}
\begin{equation}
\left( \frac{\epsilon}{\eta} \right)^{1/6} \, \eta < \; {\rm  few \; MeV } \, ,
\label{CMB}
\end{equation}
on the wall tension. The tension of the strings is also of order $\eta$ and values compatible with (\ref{CMB}) do not lead to any CMB limit from the strings. 

Another relevant constraint is related to  avoiding excessive emission of dilatonic quanta in supernovae. In the current model, such quanta in the vacuum $n=1$ have a squared  mass $m_\phi^2 = 9 \, \epsilon \, \eta$ (these are the light excitations described by (\ref{lag-dom})). If this mass is smaller than the supernovae temperature, a large number of dilatonic quanta are produced in supernovae. Such quanta decay back into photons with a rate $\Gamma \sim \xi_F^2 m_\phi^3 / M_*^2 \sim \xi_F^2 \,  \epsilon^{3/2} \, \eta^{3/2} / M_*^2$. If the corresponding decay length is smaller than the size of the supernova core, we do not have energy loss into such quanta \cite{opu}. This occurs for 
\begin{equation}
\epsilon   \ga {\rm O } \left( 10^{-2} \right) {\rm MeV} \, \left( \frac{10^{-6}}{\xi_F} \right) \, \frac{M_*}{\eta} \, ,
\label{SN}
\end{equation}
(for typical supernovae temperatures of order $T \sim 30 \, {\rm MeV}$). If we make the natural choice $\eta = {\rm O } \left( M_* \right)$, we see that the two conditions (\ref{CMB}) and (\ref{SN}) indeed allow the hierarchical choice $\epsilon \ll \eta$.

Next, we need to discuss the modification of the potential due to the interaction of $\Phi$ with matter.
As discussed in \cite{opu}, the interaction generates an additional term in the potential
\begin{equation}
\Delta V = \xi_N \, \frac{{\rm Re } \, \Phi}{M_*} \, \rho_b \, ,
\end{equation}
where $ \rho_b  $ is the energy density of baryons in the universe, and $\xi_N$ is the constant controlling the coupling of the dilaton to nucleons  in (\ref{B-exp}). As in \cite{opu},  we assume that this coupling is generated radiatively starting from the coupling of the dilaton  to photons \cite{op1}, and we disregard the contribution of neutrons. This gives $\xi_N \simeq -0.0007 \, \xi_F$ \cite{gl}. We then find, for the three vacua (\ref{3min})
\begin{eqnarray}
&&
\Delta V_1 = 0 \;\;,\;\; 
\Delta V_2 = + \Delta_b \;\;,\;\;
\Delta V_3 = - \Delta_b \nonumber\\
&&
\Delta_b \simeq 4 \times 10^{-10} \left( \frac{\xi_F}{10^{-6}} \right) \frac{\eta}{M_*} \, \rho_b^{(0)} \left( 1 + z \right)^3 \, ,
\label{dv-matter}
\end{eqnarray}
where $\rho_b^{(0)} \simeq 1.8 \times 10^{-48} \, {\rm GeV}^4$ is the current value of the baryon energy density.

The coupling to baryons breaks the degeneracy between the energies of any two domains separated by a wall. This generates motion of a wall towards the region of greater potential energy, however,  the motion is slowed down
 by Hubble friction. The velocity $v$ of the wall is described by \cite{opu}
\begin{equation}
\frac{d}{d t} \left( R \, \gamma \, v \right) = R \, \frac{\Delta V}{{\rm tension}_{\rm wall}} \, ,
\label{motion}
\end{equation}
where $R$ is the radius of the universe and $\gamma = 1/ \sqrt{1-v^2}$. For the single wall model studied in  \cite{opu}, this resulted in a current wall velocity, $v_0 \simeq 0.004$. The only significant difference between the single and the double wall model is the parametric dependence on $\epsilon \ll \eta$ of the wall tension. This results in an increase of ${\rm O } \left( \sqrt{\frac{\eta}{\epsilon}} \right)$ for the value of  $\gamma \, v$ in the model considered here with respect to the one in \cite{opu}.
This means that for the  model with two walls, we estimate $v_0 \simeq 0.004 \, \sqrt{\frac{\eta}{\epsilon}} $ as long as the current  motion is non-relativistic. The two bounds  (\ref{CMB}) and (\ref{SN}) suggest a hierarchy $\epsilon / \eta \sim 10^{-2}$, in which case it is also safe to disregard the motion for the two-wall models.

A second potential worry associated with (\ref{dv-matter}) is the tunneling of the domains $n=1,2$ into the domain of lowest energy, $n=3$. The tunneling rate is suppressed by
\cite{Kobzarev:1974cp,Coleman:1977py}
\begin{equation}
\Gamma \propto {\rm exp } \left( - \frac{27 \pi^2}{2} \, \frac{ T^4}{\Delta V^3} \right) \sim
{\rm exp} \left( - \frac{10^{135}}{\left( 1 + z \right)^9} \right) \, ,
\end{equation}
where $T$ is the tension evaluated in (\ref{tension}) and $\Delta V$ is the potential difference (\ref{dv-matter}). In the numerical evaluation, we have set $\epsilon = 10^{-2} \, \eta ,\, \eta = M_* = {\rm MeV} ,\, \xi_F = 10^{-6} \,$. The wall forms at $z \sim 10^{9}$. We see that the tunneling rate is always negligible.

\section{Data used} \label{data}

The available data on possible variations in $\alpha$ come from two sources:
Keck data \cite{Webb} containing 140 absorption systems and VLT data \cite{newwebb}
containing 153 systems.  

Concerning the VLT measurements, we use the results appearing in Table A1 of the second reference in \cite{newwebb}. Consistently with what done in  \cite{newwebb}, we discard the fourth absorber in the list as an outlier. We use the last column of that table for the values of $\Delta \alpha / \alpha$ for each of the absorbers. In several fits using
these data, ref.  \cite{newwebb} added in quadrature a common ``random error'' $\sigma_{\rm ran} = 0.905 \times 10^{-5}$ to the value of $\sigma$ given in the table for each absorber. The motivation given in  \cite{newwebb} is that the VLT data are too scattered, and therefore the original error given in the table must be underestimating the true error.  Ref.  \cite{newwebb} actually uses slightly different values for $\sigma_{\rm ran}$ for fits of the data to different functional forms. The mathematical procedure leading to the values used is described in their Section 3.5.3. This procedure ultimately allows the data to agree with the fitting model (the random error is determined by requiring that a subset of the data - once the biggest outliers are excluded - has a reduced $\chi^2$ equal to one for the model under which the data are being fit). This is certainly a necessary condition for a meaningful model comparison, if one wants to claim that the better model fits the data. However, we stress that the procedure is arbitrary to a large extent, and that ref.  \cite{newwebb} did not identify the physical source of this additional error. We therefore regard as unmotivated any p-value obtained for the fits when $\sigma_{\rm ran}$ is added; we agree that it is nonetheless interesting to compare the significance of different fits to the data, in the hope that the physical source of the error is eventually found, and that this error is not correlated with the spatial position of the observers. Concerning this last comment, we stress   that, for $\sigma_{\rm ran} = 0$, all of the models considered in  \cite{newwebb} and in this paper are a very bad fit to the data, and therefore we cannot rule out that any claimed spatial dependence is spurious. In our data analysis, we compute the $p$-value of the fits with and without adding  $\sigma_{\rm ran}$, in order to see how the significance of the dipole and wall fits is affected by this random error.

Concerning the Keck/Hires data, we use the same data studied in \cite{newwebb}. Such data were
analyzed in \cite{Murphy:2003mi,Murphy:2009az}, and  can be found in the On-line Data Table 1 linked to \cite{link}. This table contains $143$ absorbers, $27$ of which are denoted as a ``high-contrast sample''; such absorbers are at redshift $z > 1.8$ and both strong and weak transition lines had to be fit to extract the value of $\alpha$ associated to them. According to  \cite{Murphy:2003mi} this is the origin of the excess scatter present in these data, which is not adequately reproduced by the calculated statistical error. To account for this, ref.  \cite{Murphy:2003mi} added in quadrature the common error $\sigma = 1.75 \times 10^{-5}$ to the individual error of each high-contrast absorber. The value of this common error was chosen in  \cite{Murphy:2003mi} so that, once the total error is used, a common mean fits these $27$ absorbers with a reduced $\chi^2$ equal to one. Ref.  \cite{newwebb} used different values for this common error according to different fits of the data, analogously to what done for the VLT data. In our study, we compare the significance of different models both with and without the additional error,  $\sigma = 1.63 \times 10^{-5}$, which is the value used in  \cite{newwebb} for most of the dipole fits. We stress that this error is added only to the  $27$ ``high-contrast'' absorbers.
According to ref.  \cite{Griest:2009rv}, the original values of $\alpha$ studied in  \cite{Murphy:2003mi}  are affected by an incorrect wavelength calibration. 
Ref. \cite{Murphy:2009az} estimated the shift induced by the miscalibration on each individual absorber. In most cases, the quoted values are smaller than the statistical error on each measurement, and ref.  \cite{Murphy:2009az} concluded that this effect does not significantly impact the data analysis. This recalibration was disregarded in the analysis of 
 \cite{newwebb}. Additionally, ref.  \cite{newwebb} indicated that two absorbers in the table in \cite{link}, and included in the previous analysis, are affected by overly large calibration problems, and should be disregarded. One additional absorber is also identified as an outlier in the analysis of  \cite{newwebb}, and disregarded. To compare our results with those of  \cite{newwebb}, we also disregard 
the recalibration and  these three absorbers in our study.

\section{Wall fitting procedure}  \label{fit-wall} 

Let us first discuss our fitting procedure for the single domain  wall model. The wall separates the dilatonic vacuum we live in (characterized by the value of $\alpha$ measured on Earth) from the other vacuum (characterized by a different value of $\alpha$).  We need to establish an algorithm that, for any wall configuration, indicates whether a given absorber is in our vacuum, or in the other one.

As discussed above, the wall can be assumed to be planar and static today. This means that 
one can choose coordinates so that the wall is at a constant comoving cartesian coordinate, say $x=x_{\rm wall}$. The wall has a point which is closest to us. Let us denote this point by $P$. The position of $P$ is characterized by the two  angular polar coordinates $\left( \theta_* ,\, \phi_* \right)$ and by the redshift $z_*$. An additional parameter of the model is the value of $\alpha$ in the other vacuum. These four parameters completely specify the wall model for the purpose of data fitting.

\begin{figure}
\centerline{
\includegraphics[width=0.5\textwidth]{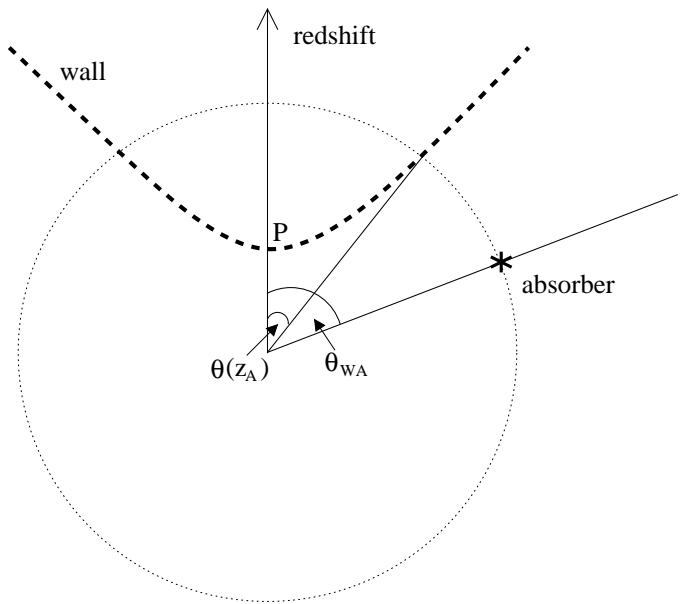}
}
\caption{Illustration of how we determine whether an absorber is in our vacuum, or in a different one. Our location is placed at the origin, and increasing radial distance correspond to increasing redshift. The Figure  shows the wall as it appears on our sky and  an absorber at redshift $z_A$. This redshift is shown as the dotted circle. The angle $\theta \left( z_A \right)$ is the angle at which we see the wall at that redshift; the angle $\theta_{WA}$ is instead the angle between the center of the wall and the absorber. The case shown has $\theta_{WA} > \theta \left( z_A \right)$ and the absorber is in our vacuum. 
}
\label{fig:wall-z-th}
\end{figure}

The dashed line shown  in Figure \ref{fig:wall-z-th} illustrates how the wall appears on our sky. In the Figure, the coordinates  have been chosen so that the origin corresponds to our location, while points  at greater radial distance from the origin correspond to points at greater redshift on our sky (we did not attempt to show this relation to scale in the Figure). The coordinates in the Figure have also been chosen so that the point $P$ lies on the positive vertical axis, $\theta_* = 0$.   The wall is described by a function $\theta \left( z \right)$, that relates the redshift $z$ of a generic point on the wall to the angle $\theta$ formed by the line of sight to this point (as seen by us), and the line of sight to the point $P$. To obtain this function, we first note that the radial comoving coordinate of this generic point on the wall is related to this angle by
\begin{equation}
r = \frac{r_*}{\cos \theta} \, ,
\label{wall-rth}
\end{equation}
 where $r_*$ is the radial comoving coordinate of $P$. We then recall the relations $a \, d r = d t = d a / \left( a H \right)$, where $a$ is the scale factor, and $H$ the Hubble rate. We insert the expression of the scale factor in terms of redshift, $a = a_0 / \left( 1 + z \right)$ in this differential relation ($a_0$ denotes the value of the scale factor today) and we integrate the resulting equation to obtain a relation between $r$ and $z$. We finally insert this result into (\ref{wall-rth}), to obtain
\begin{equation}
\cos \theta \left( z \right) = \frac{\int_0^{z_*} \frac{d z'}{H \left( z' \right)}}{\int_0^{z} \frac{d z'}{H \left( z' \right)}} = \frac{\int_0^{z_*} \frac{d z'}{\sqrt{\Omega_{\Lambda,0} + \Omega_{m,0}  \left( 1 + z' \right)^3}}}{\int_0^{z} \frac{d z'}{\sqrt{\Omega_{\Lambda,0} + \Omega_{m,0}  \left( 1 + z' \right)^3}}} \, ,
\end{equation}
where we have restricted our attention to a flat universe containing only matter and a cosmological constant. $\Omega_{\Lambda,0} \simeq 0.728$ and $\Omega_{m,0} = 1 - \Omega_{\Lambda,0}$ denote the present fractional energies of cosmological constant and matter, respectively (the numerical value is the Maximal Likelihood value given in  \cite{Komatsu:2010fb} for a $\Lambda$CDM universe, using WMAP7, BAO, and $H_0$ data). Upon changing the integration variable $z' \rightarrow y - 1$, this relation becomes
\begin{equation}
\cos \theta \left( z \right) = \frac{{\cal F} \left( z_* \right)}{{\cal F} \left( z \right)} \;\;\;,\;\;\;
{\cal F} \left( z \right) \equiv \int_1^{1+z} \frac{d y}{\sqrt{\frac{\Omega_{\Lambda,0}}{1-\Omega_{\Lambda,0}} + y^3}} \, .
\label{theta-z}
\end{equation}

This relation correctly reproduces $\theta \left( z_* \right) = 0$ for the case in which the generic point on the wall coincides with $P$. It also indicates that the wall is seen within a maximal opening angle, which is mathematically given by $\theta \left( z = \infty \right)$ (for fitting purposes, there is no difference between considering the maximal opening angle given by $z = \infty$ and the one given by the redshift at which the wall formed, since the latter quantity is much greater than the redshifts of the absorbers). The opening angle is maximal  for $z_* = 0$, when the wall passes through our location. In this limit eq. (\ref{theta-z}) gives $\theta = \pi/2$, which correctly indicates that the wall is seen in half of the sky.

It is now straightforward to describe the algorithm that we use to determine whether an absorber is in our vacuum or in the other one. Figure \ref{fig:wall-z-th} also illustrates the position of a generic absorber. For each wall configuration, and for each absorber, we compute the angle between the line of sight of the absorber and the line of sight of $P$. This is the angle $\theta_{\rm WA}$ indicated in the Figure.
We then insert the redshift of the absorber $z_A$ into eq. (\ref{theta-z}). The corresponding quantity $\theta \left( z_A \right)$ is also  shown in the Figure, and it corresponds to the angle between the line of sight of $P$ and the line of sight of a point on the wall having the redshift $z_A$. If, as in the Figure,  $\theta_{\rm WA} > \theta \left( z_A \right)$, the absorber is seen by us outside the region delimited by the wall for that redshift. Therefore the absorber is in our vacuum. If instead   $\theta_{\rm WA} < \theta \left( z_A \right)$, the absorber is ``beyond the wall'', and therefore in the other vacuum.

In the case of two walls, one could have the situation in which (i) the walls are far apart, so that they do not intersect (at least in the region occupied by the absorbers), or (ii) they are close enough that they do intersect. In case (i), the stationary and planar approximation holds for both walls, and we readily extend to this case the procedure just outlined for the single wall case. In the case (ii), one should in principle solve the field theory equations at the intersection of the two walls to determine the precise spatial distribution of the different vacuum domains. For simplicity, we disregard the configurations of the type (ii) in our analysis. In practice, referring to Figure \ref{fig:string-wall}, we imagine that the string connecting the walls is at a comparable but slightly  greater distance relative to the Keck and VLT absorbers. Concretely, in our fits of the two-wall model, we first choose the $6$ parameters that characterized the geometry of the walls (three per wall, as described above), implicitly assuming that the walls do not intersect. We then use (in our numerical fitting program) the algorithm described above to determine whether an absorber is beyond either of the two walls. If the algorithm gives that no  absorber is beyond both walls, then the assumption that the two walls do not intersect (in the region occupied by the absorbers) is correct, and the configuration is indeed of the type (i). Otherwise, the configuration is of the type (ii), and we disregard the initial choice of the $6$ parameters in the data analysis. Therefore, when we fit the two-wall model, we are only fitting the data against a subset of possible configurations. This effectively results in the fact that the significance that we quote for the  two-wall model is a conservative figure: we cannot exclude that a configuration of the type (ii) would be a better fit to the data than those that  we probe in our analysis.

\section{Results} \label{fit-results}

We perform separate fits for the Keck measurements alone ($140$ objects), for the VLT measurements alone ($153$ objects), and for the total combined  set ($293$ objets). The data were described in Section \ref{data}. The data are given in  Equatorial J2000 coordinates; the polar angles $\theta$ and $\phi$ correspond to ``declination'' (given in degrees) and ``ascension'' (given in hours), respectively. The relation between the different units is
\begin{eqnarray}
\frac{\theta}{\rm radians} & = & \frac{\pi}{2} \left( 1 - \frac{\rm declination}{90^\circ} \right) \nonumber\\
\frac{\phi}{\rm radians} & = & \frac{\pi}{12} \,   \frac{\rm ascension}{\rm hours}  \, .
\end{eqnarray}

The data are fit against the 

\begin{itemize}

\item Standard Model: no variation of $\alpha$ with respect to that measured on Earth ($0$ free parameters);

\item a monopole template: a common constant   $m = \frac{\alpha - \alpha_0}{\alpha_0}$, where $\alpha_0$ is the terrestrial value ($1$ free parameter)

\item the dipole template (\ref{dipole}) ($4$ free parameters)

\item the one wall model of \cite{opu} ($4$ free parameters)

\item the two-wall model introduced in Section \ref{twowalls}  ($7$ free parameters)

\end{itemize}

Two parameters of the dipole fit are the dipole and monopole amplitudes $A$ and $m$ specified in (\ref{dipole}). The other two parameters, $\theta$ and $\phi$, are the angles that specify 
 the direction of the dipole in a given coordinate system (specifically, we use Equatorial J2000 coordinates  in all our fits). The angle $\theta$ should not be confused with the angle $\vartheta$ defined after eq. (\ref{dipole}). The two angles $\theta$ and $\phi$ actually point towards
 a single position on the sphere, and therefore  also specify the orientation of the dipole.   Reversing this  orientation ($\theta \rightarrow \pi - \theta \;,\; \phi \rightarrow \phi  + \pi$, if the angles are expressed in radians), and changing the sign of $A$ in (\ref{dipole}) results in the same dipole fit.

The four  parameters characterizing the one wall model are the two angular coordinates and the redshift of the  point of the wall closest to us, and the value $\Delta =  \frac{\alpha - \alpha_0}{\alpha_0}$ corresponding to the vacuum on the other side of the wall with respect to us. In the two-wall model there are three additional free parameters characterizing the position of the second wall; in principle, an eighth parameter could be immediately introduced, since one can trivially modify the model introduced in Section \ref{twowalls} to have two independent values  $\Delta_i$ beyond the two walls. We remove half of the freedom by imposing that the values of $\Delta_i$ beyond the two walls are opposite to each other. This is done to reduce the dimensionality of the parameter space, in order to facilitate the search for the best fitting parameters. Due to this choice,  changing the sign of $\Delta$ and the order of the two walls results in the same configuration. Finally, as mentioned at the end of the previous Section, we exclude from the fit the configurations for which the two walls intersect each other (at least, within the region occupied by the absorbers). This assumption is made for simplicity (otherwise we would need to study the dynamics of the intersection, or at least introduce additional parameters specifying the position of a third wall, see Figure \ref{fig:string-wall}).

For any fit, the   $\chi^2$ of any parameter choice  is computed in the standard way
\begin{equation}
\chi^2 \left[ {\rm parameters} \right] \equiv 
\sum_i \frac{\left( \frac{\Delta \alpha}{\alpha} \vert_i -  \frac{\Delta \alpha}{\alpha} \vert_{\rm parameters}
\right)^2}{\sigma_i^2}
\end{equation}
where the sum is performed over the absorbers in a given dataset, and where $\sigma_i$ is the error on the $ \frac{\Delta \alpha}{\alpha} \vert_i $ measurement; $ \frac{\Delta \alpha}{\alpha} \vert_{\rm parameters} $ is instead the theoretical value for that parameter choice. The likelihood  is defined as  ${\cal L} = e^{-\chi^2/2}$. 

The dipole and walls fits are performed with a Markov chain. Each point in the chain corresponds to a point in the $n-$dimensional space of the $n$ parameters that are being fit. The first point in the chain is chosen at random. The chain is then characterized by the algorithm that allows one to add the 
$n+1-$th point to it starting from the $n-$th point. The addition is performed by identifying a candidate point and by accepting it with a given probability. The identification is performed by doing   a small random step in  parameter space starting from the $n-$th point in the chain. The candidate point is accepted with the probability 
\begin{equation}
{\rm probability \; accepting \; point } \; n+1 = {\rm Min} \; \left[ \frac{{\cal L}_{n+1}}{{\cal L}_n} ,\, 1 \right] \, ,
\end{equation}
where ${\cal L}_n$ is the likelihood of the $n-$th point, and analogously for $n+1$. We note that a candidate point is always accepted if it has a better likelihood than the point that was last accepted in the chain; otherwise, the probability decreases in proportion to how worse the likelihood of the candidate point is. A long chain then behaves as a grid which is more dense in regions with higher likelihood.

\begin{figure}
\centerline{
\includegraphics[width=0.35\textwidth,angle=-90]{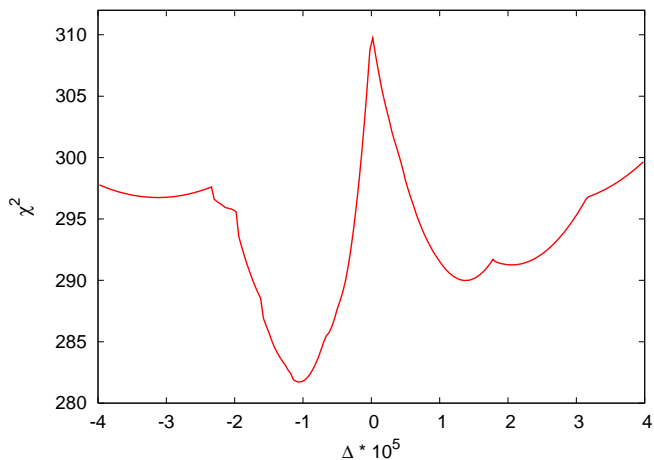}
}
\caption{$\chi^2$ vs. $\Delta$ for the one-wall model fit of the total data; see the main text for details.}
\label{fig:1w-chi2-delta}
\end{figure}

\begin{figure}
\centerline{
\includegraphics[width=0.35\textwidth,angle=-90]{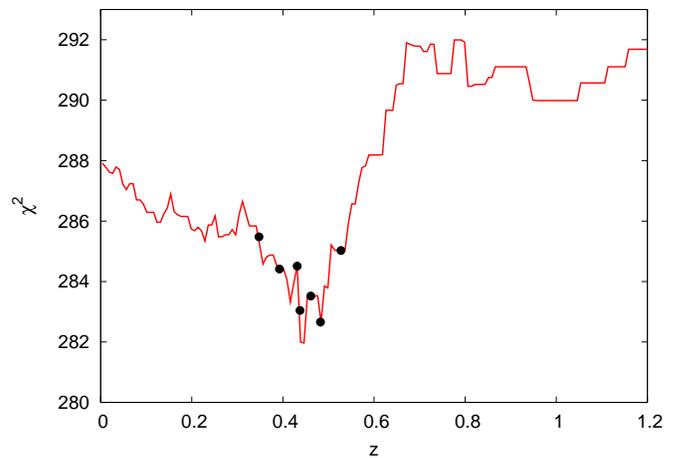}
}
\caption{$\chi^2$ vs. redshift of the closest point on the wall, for the one-wall model fit of the total data.}
\label{fig:1w-chi2-z}
\end{figure}

\begin{figure}
\centerline{
\includegraphics[width=0.35\textwidth,angle=-90]{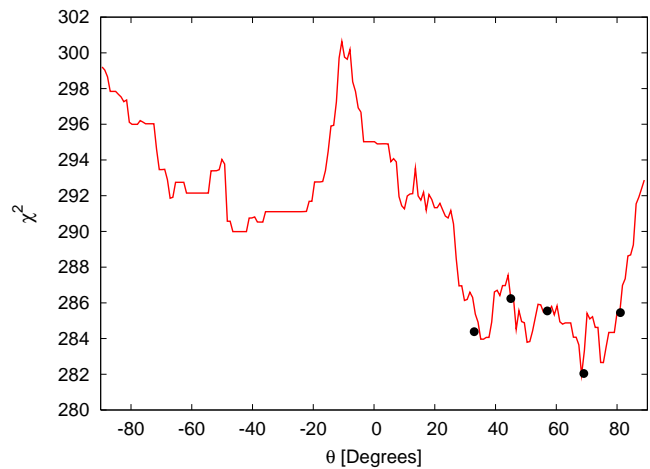}
}
\caption{$\chi^2$ vs. declination of the closest point on the wall, for the one-wall model fit of the total data.}
\label{fig:1w-chi2-th}
\end{figure}

\begin{figure}
\centerline{
\includegraphics[width=0.35\textwidth,angle=-90]{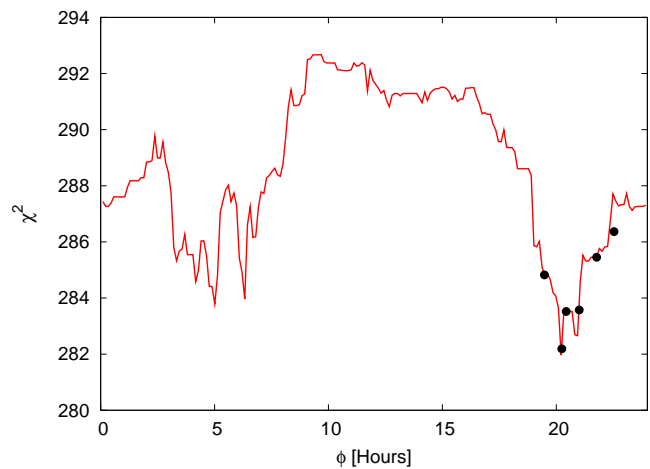}
}
\caption{$\chi^2$ vs. ascension of the closest point on the wall, for the one-wall model fit of the total data.}
\label{fig:1w-chi2-phi}
\end{figure}

The point in the chain with smallest $\chi^2$ is our solution for the parameters that best describe the data within a given fit. To find the confidence intervals around this best point, we vary the value of one parameter at a time, and marginalize over the other parameters (namely, the other parameters are free to vary, until the configuration with the minimal $\chi^2$ is found). For the dipole fits, the $\chi^2$ distribution is approximately Gaussian around the best fit point. Therefore, we determine the (approximate) $1 \sigma$ confidence level by varying each parameter until the $\chi^2$ (with the other parameters marginalized over) increases by $1$ with respect to the best fit point. For the wall fits, this was possible only for the parameter  $\Delta$
and because the distribution does not resemble a Gaussian, the 68\% CLs are determined
directly from the the likelihood function. For the other parameters, the $\chi^2$ behavior is too irregular. We show the $\chi^2$ distribution for some of the fits in Figures \ref{fig:1w-chi2-delta},  \ref{fig:1w-chi2-z},  \ref{fig:1w-chi2-th}, and \ref{fig:1w-chi2-phi}.

The results of our analysis are summarized in a number of Tables and Figures. In the Tables,  we give the $\chi^2$ and the corresponding $p-$values of a number of fits to the data. We recall that the $p$ value $p=\Gamma \left( \frac{\nu}{2} ,\, \frac{\chi^2}{2} \right) / \Gamma \left( \frac{\nu}{2} \right)$
(where $\nu$ represents the number of degrees of freedom) indicates the probability that a set of fictitious data generated from the model being studied has  a greater $\chi^2$ than the fit of the actual data. Therefore, it is a measure of how well the model  fits the data. 

The results given in Table \ref{table:norandomerr}  are derived when omitting the additional  random errors discussed in Section \ref{data}.
For the Keck data, we see a significant drop in $\chi^2$ for the monopole which was
the basis of the claim in \cite{Webb} for a temporal variation in $\alpha$. There is little motivation here
for a dipole, but the single wall solution does bring another significant drop in $\chi^2$. 
Here, adding a second wall does very little. For the one-wall solution, the $p$-value reaches
4 \%. For the VLT data, the monopole solution offers a modest improvement in $\chi^2$, 
and we see that $\chi^2$ has significant improvements as we move from the dipole
to the two-wall solution.  In each case, the $p$-value is extremely small.
For the combined data, the monopole offers almost no improvement, while the dipole
and one-wall solutions give an almost identical drop in $\chi^2$, which is further lowered in 
the two-wall solution.  In total, $\chi^2$ is lowered by 100, at the cost of seven parameters.
As we see from the very low $p$-values, none of the fits are good representations of the data.

\begin{center}
\begin{table}
\begin{tabular}{|l|l|l|l|}
\hline
Fit & Keck & VLT & Combined \\ \hline
Std. Model & $217.1 \; ( 3 \times 10^{-5} )$ & $280.2 \; (2 \times 10^{-9})$ & $497.3 \; (10^{-12} )$ \\ \hline
Monopole & $180.3 \; ( 10^{-2} )$  &  $269.9 \; (  10^{-8} )$  & $496.5 \; (10^{-12} )$ \\ \hline
Dipole & $176.5 \; (   10^{-2} )$ & $250.8 \; ( 4 \times 10^{-7} )$ & $449.6 \; ( 4 \times 10^{-9} )$ \\ \hline
One wall & $165.9 \; ( 4 \times 10^{-2} )$ & $229.5 \; ( 3 \times 10^{-5} )$ & $449.2 \; ( 5 \times 10^{-9} )$ \\ \hline
Two walls & $162.0 \; ( 4 \times 10^{-2} )$  & $209.0 \; ( 5 \times 10^{-4} )$ & $397.0 \; (   10^{-5} )$ \\ \hline
\end{tabular}
\caption{$\chi^2$ (and $p-$values) without adding random errors on the VLT data or the $27$ ``high contrast'' Keck data.}
\label{table:norandomerr}
\end{table}
\end{center}

Tables \ref{table:Keck}, \ref{table:VLT}, and \ref{table:combo} present the solutions for the best parameter values, the $\chi^2$, and the $p-$value  of the various fits for the Keck data, the VLT data, and the total combined data, respectively, with the additional random uncertainty included. 
The basic behavior of $\chi^2$ described for Table \ref{table:norandomerr} is repeated when
the extra random errors are included.  Of course, overall values of $\chi^2$ are significantly lower
with correspondingly high values of $p$. We stress that the value of the extra error is determined a posteriori during the data analysis itself  \cite{Murphy:2003mi,newwebb}, and its value is chosen so that the dipole fit is a good fit to the data. 

We note that as seen in Fig. \ref{fig:1w-chi2-delta}, there are multiple local minima for the value of 
$\Delta$ corresponding to the shift in $\alpha$. For the Keck and VLT data alone, the lowest two minima have comparable values of $\chi^2$ and as such, there are 
two distinct ranges for $\Delta$ at the 68\% CL: (-1.24 -- -1.00) and (-0.97 -- -0.68) for Keck,
and (1.01 -- 1.68) and (1.78 -- 2.33) for VLT.  For the combined data, the 68\% CL falls within
a single range for $\Delta$.

\begin{center}
\begin{table}[h]
\begin{tabular}{|l|l|l|}
\hline
Keck Fits & Parameters & $\chi^2$ ($p-$value) \\ \hline
Std. Model & & $157.9 \;\; (0.14)$  \\ \hline
Monopole & $m=-0.57 \pm 0.11$ & $132.8 \;\; (0.63)$  \\ \hline
Dipole & $m=-0.47 \pm 0.15  $ &   \\ 
 & $ A = 0.41^{+33}_{-32} $ &   \\ 
 & $\theta = -47^{+49}_{-31} \;,\; \phi = 16.0_{-4.3}^{+2.6} $
& $131.0 \;\; (0.61)$  \\ \hline
One wall & $\Delta=-0.80_{-0.44}^{+0.12} \;\;,\;\; z = 0.15  $   & \\ 
 & $ \theta = 32 \;\;,\;\; \phi = 4.0 $  & $127.6 \;\; (0.68) $ \\ \hline
Two walls & $z_1 = 0.16 \;,\; \theta_1 = 33.9 \;,\; \phi_1 = 3.9$ & \\ 
 & $z_2 = 0.74 \;,\; \theta_2 = -37.6.5 \;,\; \phi_2 = 18.2 $ & \\ 
 & $\Delta = - 0.82$ & $ 127.5 \;\; (0.62) $ \\ \hline
\end{tabular}
\caption{Fits for the Keck    data, adding  random errors on $27$ ``high contrast''  data.
The quantities $m ,\, A ,\, \Delta$ are given in units of $10^{-5}$.}
\label{table:Keck}
\end{table}
\end{center}

\begin{center}
\begin{table}[h]
\begin{tabular}{|l|l|l|}
\hline
VLT Fits & Parameters & $\chi^2$ ($p-$value) \\ \hline
Std. Model & & $152.5 \;\; (0.50)$  \\ \hline
Monopole & $m=0.21 \pm 0.13$ & $149.8 \;\; (0.54)$  \\ \hline
Dipole & $m=-0.11 \pm 0.19  $ &   \\ 
 & $ A =  1.17^{+0.47}_{-0.46} $ &   \\ 
 & $\theta = -62 \pm 14 \;,\; \phi = 18.3_{-1.3}^{+1.5}  $
& $141.8 \;\; (0.65)$  \\ \hline
One wall & $\Delta=1.38_{-0.37}^{+0.95} \;\;,\;\;  z = 1.34  $   & \\ 
 & $ \theta = -37.8 \;\;,\;\; \phi = 20.1  $   & $132.7 \;\; (0.83) $ \\ \hline
Two walls & $z_1 = 1.04 \;,\; \theta_1 = -42.9 \;,\; \phi_1 = 19.3$ & \\ 
 & $z_2 = 1.50 \;,\; \theta_2 = -14.6 \;,\; \phi_2 = 8.1 $ & \\ 
 & $\Delta =  1.44$ & $ 127.2 \;\; (0.87) $ \\ \hline
\end{tabular}
\caption{Fits for the VLT data, including  random errors.}
\label{table:VLT}
\end{table}
\end{center}

\begin{center}
\begin{table}[h]
\begin{tabular}{|l|l|l|}
\hline
Comb. Fits & Parameters & $\chi^2$ ($p-$value) \\ \hline
Std. Model & & $310.4 \;\; (0.23)$  \\ \hline
Monopole & $m=-0.22 \pm 0.08$ & $303.7 \;\; (0.31)$  \\ \hline
Dipole & $m=-0.18 \pm 0.09  $ &   \\ 
 & $ A = 0.97 \pm 0.21 $ &   \\ 
 & $\theta = -61 \pm 10 \;,\; \phi = 17.3_{-1.1}^{+1.0}  $
& $280.6 \;\; (0.63)$  \\ \hline
One wall & $\Delta=-1.06_{-0.22}^{+0.24} \;\;,\;\; z = 0.45 \;\; $   & \\ 
 & $ \theta = 68.3 \;\;,\;\; \phi = 20.2 \;\; $   & $282.0 \;\; (0.60) $ \\ \hline
Two walls & $z_1 = 0.44 \;,\; \theta_1 = 67.9 \;,\; \phi_1 = 20.1$ & \\ 
 & $z_2 = 1.02 \;,\; \theta_2 = -39.4 \;,\; \phi_2 = 19.2 $ & \\ 
 & $\Delta = - 1.12$ & $ 263.2 \;\; (0.83) $ \\ \hline
\end{tabular}
\caption{Fits for the Keck and the VLT data, including random errors on $27$ ``high contrast'' Keck data, and on all of the VLT data.}
\label{table:combo}
\end{table}
\end{center}

We see that the dipole or wall solutions offer
little improvement for the Keck data alone. The single wall gives a large improvement 
for the VLT data. For the combined data, we see again that the single wall and dipole 
fits are clearly better than the Standard Model or the monopole fit, while the two-wall fit offers 
the best solution to the combined data.

Figures \ref{fig:1w-chi2-delta},  \ref{fig:1w-chi2-z},  \ref{fig:1w-chi2-th},  \ref{fig:1w-chi2-phi}, correspond to the one-wall fit of the combined data 
(with the additional random uncertainty included). In each Figure
we show the $\chi^2$ obtained by keeping one parameter  fixed  (to the value shown on the horizontal axis of the Figure) and by varying the remaining $3$ parameters until the minimum $\chi^2$ is found (i.e., by marginalizing over the remaining $3$ parameters). For each Figure, $200$ values of the fixed parameter plus the resulting $\chi^2$ are connected by a solid line. For each value, the marginalization procedure is done with a Markov chain of $500,000$ points. The large dots visible in three of the Figures are obtained with a chain of $20$ million points, and verify that the shorter chain used to obtain the solid line is sufficient for the marginalization.

\begin{figure}[h!]
\centerline{
\includegraphics[width=0.45\textwidth]{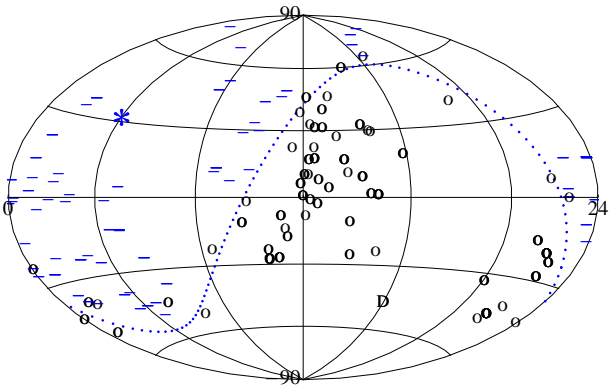}
}
\caption{Keck absorbers (marked according to which vacuum they are in) and wall configuration for the single wall best fit given in Table \ref{table:Keck}. The asterisk indicates the position of the point on the wall closest to us; the dashed line indicates the boundary of the wall. The letter D marks the direction of the dipole, oriented where the dipole amplitude is greatest.
}
\label{fig:sky-1w-Keck}
\end{figure}

In Figure \ref{fig:sky-1w-Keck}, we show the position of the Keck absorbers on our sky and the wall configuration for the single wall best fit given in Table \ref{table:Keck}. Each absorber is marked by a zero sign if it is in our vacuum (since $\Delta \alpha = 0$ in this case), of by a minus sign if it is in the other vacuum (since $\Delta \alpha < 0$ in this case).  The location of the point on the wall which is closest to us is marked with an asterisk. The wall boundary, shown with a dashed line in the Figure, corresponds to the (mathematical) locations at which the wall would be seen at infinite redshift on our sky. The wall intersects any line of sight pointing in the region limited by that boundary (and including the asterisk); the redshift of the intersection is an increasing function of the angle between that line of sight and the line of sight of the nearest point of the wall (this corresponds to the angle denoted by $\theta \left( z_A \right)$ in Figure \ref{fig:wall-z-th}). An absorber in that portion of the sky  can either be in our vacuum if it is  ``in front of'' the wall (namely if its redshift is smaller than the redshift at which the wall intersects the line of sight of that absorber), or in the other vacuum if is is  ``behind'' the wall. All the absorbers in the other portion of the sky (the one delimited by the dashed line boundary, and not containing the asterisk) are instead  in our vacuum. Also shown in  Figure \ref{fig:sky-1w-Keck} is
the position of the dipole solution, marked by a `D',  which is well aligned (perpendicular) to the wall.

\begin{figure}[h]
\centerline{
\includegraphics[width=0.45\textwidth]{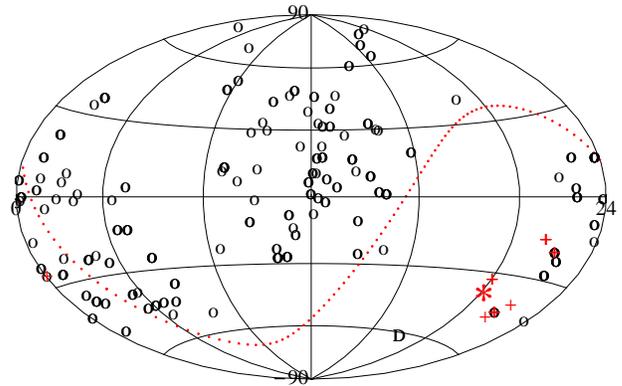}
}
\caption{VLT absorbers and wall configuration for the single wall best fit given in Table  \ref{table:VLT}. 
}
\label{fig:sky-1w-VLT}
\end{figure}

\begin{figure}[h]
\centerline{
\includegraphics[width=0.45\textwidth]{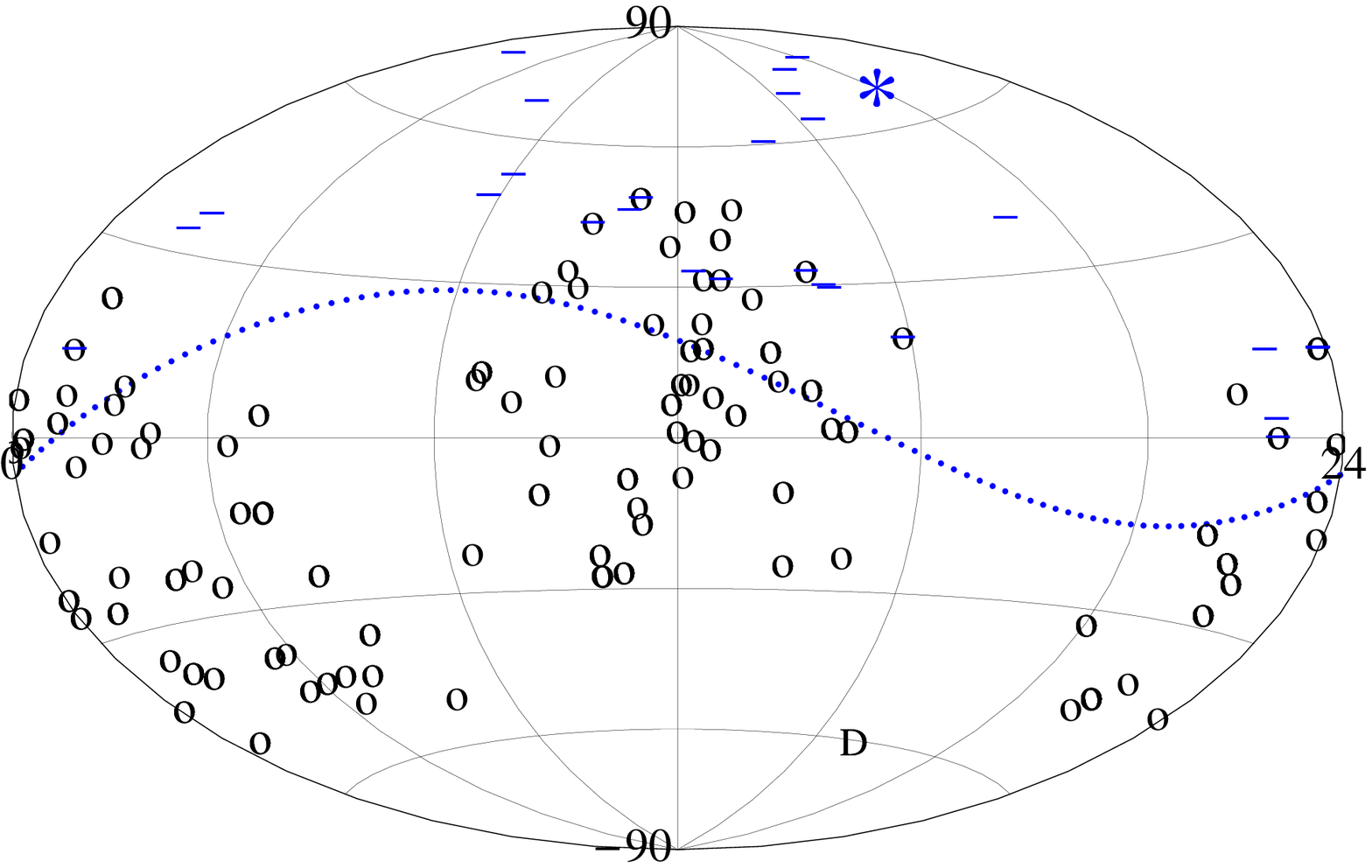}
}
\caption{Total absorbers sample and wall configuration  for the single wall best fit given in Table  \ref{table:combo}. 
}
\label{fig:sky-1w-TOT}
\end{figure}

\begin{figure}[h!]
\centerline{
\includegraphics[width=0.45\textwidth]{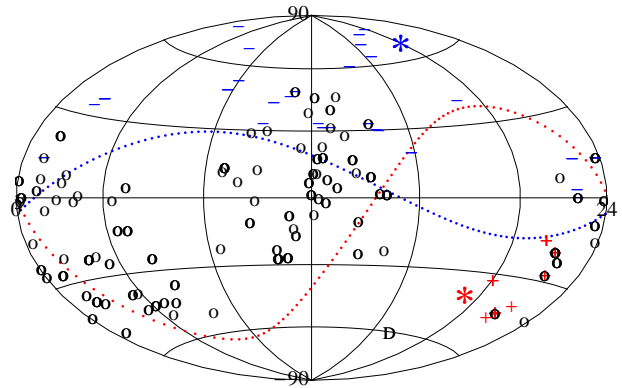}
}
\caption{Total absorbers sample and wall configuration  for the double wall best fit given in Table  \ref{table:combo}. 
}
\label{fig:sky-2w-TOT}
\end{figure}

The VLT absorbers and the corresponding single wall best fit (Table  \ref{table:VLT}) are shown in Figure \ref{fig:sky-1w-VLT}. The absorbers in the other vacuum are marked with a plus sign in this Figure, since $\Delta \alpha > 0$ in that vacuum. 
The wall solution in this case is found in a different region of the sky, though it too is well aligned
with the dipole (which moved very little).
The combined sample and the corresponding single wall best fit (Table \ref{table:combo}) are shown instead in Figure \ref{fig:sky-1w-TOT}. 
In this case, the single wall attempts to interpolate between the two previous solution
and is no longer aligned with the dipole.
Finally, Figure \ref{fig:sky-2w-TOT} shows how the combined sample separates into three vacua for the two-wall best fit  (Table \ref{table:combo}). We note the presence on the sky of a region where the two wall solutions intersect. The actual wall solution should be modified in that region so to properly account for this (we expect the presence of a third wall, see Figure \ref{fig:string-wall}). However, all absorbers  in that portion of the sky have a redshift smaller than the intersection (in fact, they have a redshift smaller than the one corresponding to the second wall). Therefore, the more precise solution is not needed for the fit.

\section{Spatial dependence} \label{fit-significance}

We divide this Section in two parts. In the first part we study the significance of the fits for which    $\alpha$ exhibits spatial dependence. In the second part, we further discuss the spatial orientation of the different solutions.

\subsection{Significance of the fits}

For this discussion, we assume that the extra random errors included in \cite{Murphy:2003mi,newwebb} are a valid estimate of the actual error
in the data.  Here, we pose a well defined question of whether  ``model $2$'' (spatial variation), which includes ``model $1$'' (no spatial variation) as a sub-case,  is a significantly better fit to the data. The statistical significance of the improvement is obtained by comparing  $F = \frac{\chi_1^2-\chi_2^2}{p_1-p_2} / \frac{\chi_2^2}{n-p_2}$ (where $\chi_i^2$ and $p_i$ are the $\chi^2$ and the number of parameters  of the $i-$th model, respectively, while  $n$ is the total number of observations) of the actual data with that of a random set of data. The values reported in Table \ref{table:statsig-SM} are the 
probabilities that fitting a random set of data with the model indicated in the row results in a greater 
$\chi^2$ improvement over the Standard Model than the improvement obtained  with the actual data. 
For example, we see that for the VLT data, there is a 10\% chance that random data would yield
a comparable improvement in $\chi^2$ relative to that of the monopole.

\begin{center}
\begin{table}
\begin{tabular}{|l|l|l|l|}
\hline
Statistical preference of & Keck & VLT & Total \\ 
 Standard Model over : &  &  &  \\ \hline
Monopole  & $10^{-6}$ &  $0.10$ & $0.012$ \\ \hline
Dipole  & $4 \times 10^{-5}$ &  $0.03$ & $7 \times 10^{-6}$ \\ \hline
One wall  & $7 \times 10^{-6}$ & $3 \times 10^{-4}$  & $1.4 \times 10^{-5}$  \\ \hline
Two walls  & $1.5 \times 10^{-4}$ &  $3 \times 10^{-4}$ & $4 \times 10^{-8}$ \\ \hline
\end{tabular}
\caption{Significance of the various fits  compared with that of the Standard Model $\frac{\Delta \alpha}{\alpha} = 0$. The smaller a number in the Table, the more the model indicated in the corresponding row is statistically significant over the Standard Model. Random errors are added to the data as indicated in Section \ref{data}.}
\label{table:statsig-SM}
\end{table}
\end{center}

 Table \ref{table:statsig-monopole} shows the analogous comparison of the dipole template and the wall models against the  monopole template. While Ref. \cite{newwebb} discussed a number of uncertainties that could affect the value for $\alpha$ obtained for each absorber, none of these uncertainties is correlated with the spatial position of the absorber. Therefore the uncertainties studied in  \cite{newwebb} may result in at most a bias in the average value of $\Delta \alpha$ that one obtains from the data. Therefore, comparing the dipole and wall fits against a non-vanishing monopole term allows one to study the significance for a spatial variation of $\alpha$ including the possible systematic errors discussed in  \cite{newwebb}.

\begin{center}
\begin{table}
\begin{tabular}{|l|l|l|l|}
\hline
Statistical preference of & Keck & VLT & Total \\ 
 monopole over : &  &  &  \\ \hline
Dipole  & $0.60$ &  $0.04$ & $4 \times 10^{-5}$ \\ \hline
One wall  & $0.14$ & $4 \times 10^{-4}$  & $8 \times 10^{-5}$  \\ \hline
Two walls  & $0.48$ &  $5 \times 10^{-4}$ & $3 \times 10^{-7}$ \\ \hline
\end{tabular}
\caption{Significance of the various fits with that of a monopole template, $\frac{\Delta \alpha}{\alpha} = m$. Random errors are added to the data as indicated in Section \ref{data}.}
\label{table:statsig-monopole}
\end{table}
\end{center}

Most of the results presented in these two tables already emerged in the discussions done in the previous Section. The significance for the monopole over the Standard Model (which, as we said, could be interpreted as a signal of an overall systematic error in the data) is very strong for the Keck absorbers, and less marked for the VLT measurements. The Keck data do not show any evidence for a dipole modulation being preferred over monopole, while a stronger significance emerges for the one-wall model. The significance for the dipole is increased in the VLT data; however, the one-wall solution has a far greater significance than that of the dipole. The individual VLT and (particularly) Keck measurements do not support the two-wall model over the one-wall model. However, a very strong significance for the two-wall model appears once all the data are included. In this case, the two-wall model is a significantly better fit than the dipole template and the single-wall model, even accounting for the fact that it has three more parameters.

\subsection{Spatial orientation}

Finally, let us further study and compare  the spatial  orientations of the various fits. We confirm      that the Keck and VLT dipole orientations are consistent with each other \cite{newwebb}. The orientation of the one-wall fit of the  Keck data is also in excellent agreement  with that of the dipole fits as seen in Figure \ref{fig:sky-1w-Keck} (in the wall fits, the orientation refers to the  line of sight to the point on the wall closest to us). This is not the case for the   VLT data; in this case the  dipole and the one-wall orientations are only consistent with each other at the  $\sim 2 \sigma$ level (see Figure \ref{fig:sky-1w-VLT} and Table \ref{table:VLT}). More importantly, the domain structures indicated by the Keck and the VLT data are incompatible with each other, as they imply an opposite sign for  $\Delta$ on the other side of their respective walls. The inconsistency is also noted by the fact that the best one-wall fit of the combined sample is oriented in a different direction with respect to the two separate samples (see Figure \ref{fig:sky-1w-TOT}). 

The separate analyses are indeed rather suggestive of the presence of  two different domain walls. This is confirmed by the two-wall fits. Neither separate sample provides evidence for the two-wall fit:  each separate sample only covers a limited portion of the sky, with the Keck data more concentrated in the northern galactic hemisphere, and VLT data concentrating on more southern galactic coordinates (the different positions of the absorbers are clear from comparing Figures \ref{fig:sky-1w-Keck} and \ref{fig:sky-1w-VLT}; the position of the galactic disk in these coordinates can be seen in Figure 30 of \cite{newwebb}). The evidence  for the two domain walls only emerges when the two samples are combined.~\footnote{The non uniform distribution of the absorbers on our sky appears to be the origin of the non-gaussian $\chi^2$ behavior observed in Figures \ref{fig:1w-chi2-th} and  \ref{fig:1w-chi2-phi}. The sky coverage is not  sufficiently complete and uniform for the central limit theorem to apply.} We remark that, for the total sample of data, the double-wall fit is much more significant than either the dipole or the single-wall fits.

In comparing the different fits, one should not be misguided by the expectation that the one-wall model should always be well modeled by a dipole. This is the case only if the wall is very close to us, so that most (if not all) of the absorbers in the direction of the wall are indeed beyond it. Only in this case, the wall model assigns a nonvanishing $\Delta$ to those absorbers, and one would find that, approximately, the value of $\alpha$ of the absorbers is only function of the hemisphere in which they lie, which would indeed be well approximated by a dipole; in the opposite case, we note that  a wall further than all the absorbers provides $\Delta \alpha = 0$ for all the absorbers, exactly like the Standard Model. Therefore, we expect that a single model can be well approximated by a dipole only if the redshift of the wall is sufficiently small. This is confirmed by the data: only for the Keck sample the best dipole fit  is a good approximation to the best single-wall fit; it is not a coincidence that the value of $z_{\rm wall}$ emerging from the Keck sample is smaller than the one emerging from the VLT and from the combined samples. In this regard, the fact that the dipole and the single wall best fits of the combined data have a comparable $\chi^2$ appears to be just a coincidence.

\section{Summary} \label{summary}

In this work, we examined models which could account for a spatial variation in the electromagnetic fine structure constant from Keck  and VLT  observations. We used data  derived  by \cite{Webb, newwebb} using the many-multiplet method on each absorption spectrum. We fit this set of data against one of the dipole templates studied in  \cite{newwebb}, against the dilatonic domain wall model of \cite{opu}, and against a simple generalization of this model containing two walls. It is first of all worth stressing that all these models are very bad fits to the data (see Table \ref{table:norandomerr}) if one takes 
only the statistical error provided by the many-multiplet procedure as the error in the measurements. To compensate for this,
ref.  \cite{Webb} added a common error (in quadrature) to $27$ of the Keck absorbers, and ref. \cite{newwebb} added a common random error to all of the VLT measurements. The value of the added errors are chosen such that a monopole (in the Keck case) and a dipole (in the VLT case) are good fits to the data (we described the procedure in Section \ref{data}). For the Keck data, a case was made for adding such an error on $27$ objects, since different lines (with respect to the other absorbers) are used for deducing the value of $\alpha$ in these absorbers. An analogous justification was not given for the VLT data. Given the profound consequences that a variation of $\alpha$ would imply, it would certainly be preferable to understand and quantify this error before embarking on an analysis of the data.

Even if this problem precludes us from firmly inferring a spatial dependence from the data, it is nonetheless interesting to analyze the data under the same assumptions made in  \cite{Webb,newwebb}, to understand whether the dipole template considered in those works is robust.
Finding for example  that some  different fit describes the data better  could help in identifying possible systematic origins of the spatial variation. Furthermore, no underlying physical model was presented
in \cite{Webb,newwebb} for a dipole variation of $\alpha$. The simplest theoretical way to achieve a variation of $\alpha$ is to assume that $\alpha$ is controlled by the vacuum expectation value of a field; in string theory models this is the dilaton field. If the potential of the dilaton admits multiple minima, one may imagine a situation in which different dilatonic  domains exist in the universe, characterized by different values of $\alpha$. A variation of $\alpha$ on large spatial scales would result if  one of these domains is at a distance comparable to the typical distances of the Keck and VLT
absorbers \cite{opu}. It is interesting to see how this more physically motivated model compares against
a dipole template.

As the dipole and the wall fits  have the same number of parameters (4), an immediate comparison is obtained by comparing the resulting $\chi^2$ for each model. The wall  performs better when the Keck and VLT samples are analyzed separately. However, the Keck and the VLT best dipoles are consistent with each other,
while the  wall fits are not (the Keck wall has $\Delta < 0$ for the other vacuum, while the
VLT wall has $\Delta > 0$). As a consequence, the combined data set are fit (marginally) better by a dipole than by a single wall; the difference is however of only $\delta \chi^2 \simeq 1.4$ over $293$ objects in the sample, so that the two fits are nearly equivalent. The fact that the two samples  were best fit by two different walls, and the fact that the Keck and the VLT absorbers - with a few exceptions -   cover   different portions of the sky (so that a second wall could affect a largely different sample), prompted us to investigate the total dataset with two walls.  This turns out to be the fit with the highest significance, even if it has $3$ more parameters than either the single wall and the dipole fits.
With the extra random errors added, the statistical significance of the Standard Model over the latter fits is $ {\rm O} \left( 10^{-5} \right)$, while the statistical significance of the  Standard Model fit over the two-wall is  $ {\rm O} \left( 10^{-8} \right)$. Even when the extra errors are not included, the two-wall model is certainly the best fit of the data. The addition of its $7$ parameters allow the $\chi^2$ to decrease by 100 with respect to the Standard Model. When the random errors are
included, the drop in $\chi^2$ is 47. 

In conclusion, if the procedure adopted in  \cite{Webb,newwebb} for the extra errors turns out to be a good measure of the total errors, it is clear that the two-wall model gives a viable, statistically significant, physical model for the spatial dependence of the data. The model that we have proposed in  Section  \ref{twowalls} to account for the multiple walls undergoes  two stages of symmetry braking as the temperature of the universe decreases. Cosmic strings form at the first stage (with a tension too smaller to be of any cosmological relevance), and multiple domain walls stream off each string at the second breaking (as discussed in \cite{opu}, astrophysical and cosmological limits would force the tension of the wall to be of $\sim {\rm O } \left( {\rm MeV} \right)$). Therefore, the physical statement associated with the two wall model is that we would happen to live cosmologically close to one of the strings formed at the first breaking. On the other hand, it is also worth noting that, if  the added random errors are correct, the Standard Model itself is no longer a terrible fit to the data. Its $p-$value for the total data set is $23\%$. The significances of the dipole and wall models over the Standard Model quoted in the previous paragraph are purely statistical ones, and refer to an unbiased model comparison. It is unclear whether we should really be unbiased when comparing the Standard Model with a template with no underlying model, or even with a model that is not supported by any other measurement.

Note added:
During the completion of this work, similar results concerning the position and
statistical significance of a domain wall in relation to a spatial variation in $\alpha$
appeared \cite{newflam}.
 
\vskip.25cm
\noindent{\bf Acknowledgements:} 
We would like to thank V. Flambaum and J. Webb for an advance view of their data.
We also thank M. Murphy, C. Scarlata and M. Voloshin for useful discussions. This work  was supported in part by
DOE grant DE-FG02-94ER-40823 at the University of Minnesota.


\begin{thebibliography}{99}

\bibitem{jp-revue}
J.-P.~Uzan,
  Rev.\ Mod.\ Phys.\  {\bf 75}, 403 (2003),
  [arXiv:hep-ph/0205340];
J.-P.~Uzan,
  arXiv:1009.5514 [astro-ph.CO];
   J.-P.~Uzan,
  Gen.\ Rel.\ Grav.\  {\bf 42}, 2219 (2010)
  [arXiv:0908.2243 [astro-ph.CO]];
J.-P.~Uzan,
  Gen.\ Rel.\ Grav.\  {\bf 39}, 307 (2007)
  [arXiv:astro-ph/0605313].
  
  \bibitem{newwebb} 
  J.~K.~Webb, J.~A.~King, M.~T.~Murphy, V.~V.~Flambaum, R.~F.~Carswell and M.~B.~Bainbridge,
  Phys.\ Rev.\ Lett.\  {\bf 107}, 191101 (2011)
  [arXiv:1008.3907 [astro-ph.CO]];
   J.~A.~King, J.~K.~Webb, M.~T.~Murphy, V.~V.~Flambaum, R.~F.~Carswell, M.~B.~Bainbridge, M.~R.~Wilczynska and F.~E.~Koch,
  arXiv:1202.4758 [astro-ph.CO].
  
\bibitem{Webb}
J.~K.~Webb, {\it et al.},
 Phys.\ Rev.\ Lett.\  {\bf 82} (1999) 884
 [arXiv:astro-ph/9803165];
M.~T.~Murphy {\it et al.},
 Mon.\ Not.\ Roy.\ Astron.\ Soc.\  {\bf 327} (2001) 1208
 [arXiv:astro-ph/0012419];
J.~K.~Webb {\it et al.},
 Phys.\ Rev.\ Lett.\  {\bf 87}, 091301 (2001)
 [arXiv:astro-ph/0012539];
M.~T.~Murphy, {\it et al.},
 Mon.\ Not.\ Roy.\ Astron.\ Soc.\  {\bf 327} (2001) 1223
 [arXiv:astro-ph/0012420];
M.~T.~Murphy, {\it et al.},
 Mon.\ Not.\ Roy.\ Astron.\ Soc.\  {\bf 345}, 609 (2003)
 [arXiv:astro-ph/0306483].

\bibitem{massgrave}
G.~R.~Dvali and M.~Zaldarriaga,
 Phys.\ Rev.\ Lett.\  {\bf 88}, 091303 (2002)
 [arXiv:hep-ph/0108217];
T.~Chiba and K.~Kohri,
 Prog.\ Theor.\ Phys.\  {\bf 107} (2002) 631
 [arXiv:hep-ph/0111086];
C.~Wetterich,
 JCAP {\bf 0310}, 002 (2003);
C. Wetterich, 
 Phys.\ Lett.\ B {\bf 561}, 10 (2003)
 [hep-ph/0301261]; 
C. Wetterich, 
 [hep-ph/0302116];
L.~Anchordoqui and H.~Goldberg,
 Phys.\ Rev.\ D {\bf 68}, 083513 (2003)
 [arXiv:hep-ph/0306084];
Y.~Fujii,
  Phys.\ Lett.\  B {\bf 573}, 39 (2003)
  [arXiv:astro-ph/0307263];
E.~J.~Copeland, N.~J.~Nunes and M.~Pospelov,
 Phys.\ Rev.\ D {\bf 69}, 023501 (2004)
 [arXiv:hep-ph/0307299];
M.~Peloso and E.~Poppitz,
 Phys.\ Rev.\ D {\bf 68} (2003) 125009
 [arXiv:hep-ph/0307379];
J.-P.~Uzan,
  Phys.\ Rev.\  D {\bf 59}, 123510 (1999)
  [arXiv:gr-qc/9903004].
D.~S.~Lee, W.~Lee and K.~W.~Ng,
  Int.\ J.\ Mod.\ Phys.\  D {\bf 14}, 335 (2005)
  [arXiv:astro-ph/0309316];
W.~L.~Lee, K.~W.~Ng and D.~S.~Lee,
  Mod.\ Phys.\ Lett.\  A {\bf 19}, 1089 (2004);
N.~J.~Nunes and J.~E.~Lidsey,
  Phys.\ Rev.\  D {\bf 69}, 123511 (2004)
  [arXiv:astro-ph/0310882].
M.~Byrne and C.~Kolda,
  [arXiv:hep-ph/0402075];
P.~P.~Avelino, C.~J.~A.~Martins and J.~C.~R.~Oliveira,
  Phys.\ Rev.\  D {\bf 70}, 083506 (2004)
  [arXiv:astro-ph/0402379].
Y.~Fujii and S.~Mizuno,
  Int.\ J.\ Mod.\ Phys.\  D {\bf 14}, 677 (2005)
  [arXiv:astro-ph/0404222];
S.~Lee, K.~A.~Olive and M.~Pospelov,
  Phys.\ Rev.\  D {\bf 70}, 083503 (2004);
M.~Doran,
  JCAP {\bf 0504}, 016 (2005)
  [arXiv:astro-ph/0411606];
V.~Marra and F.~Rosati,
  JCAP {\bf 0505}, 011 (2005)
  [arXiv:astro-ph/0501515];
P.~P.~Avelino, {\it et al.},
  Phys.\ Rev.\  D {\bf 74} (2006) 083508
  [arXiv:astro-ph/0605690];
T.~Chiba,  {\it et al.},
  Phys.\ Rev.\  D {\bf 75}, 043516 (2007)
  [arXiv:hep-ph/0610027];
 S.~Lee,
  Mod.\ Phys.\ Lett.\  A {\bf 22}, 2003 (2007)
  [arXiv:astro-ph/0702063];
T.~Dent,
  JCAP {\bf 0701}, 013 (2007);
M.~E.~Mosquera,  {\it et al.},
  arXiv:0707.0661 [astro-ph].
  
\bibitem{SBM}
H.~B.~Sandvik, J.~D.~Barrow and J.~Magueijo,
  Phys.\ Rev.\ Lett.\  {\bf 88}, 031302 (2002)
  [arXiv:astro-ph/0107512];
J.~D.~Barrow, H.~B.~Sandvik and J.~Magueijo,
  Phys.\ Rev.\ D {\bf 65}, 063504 (2002)
  [arXiv:astro-ph/0109414];
J.~D.~Barrow, J.~Magueijo and H.~B.~Sandvik,
  Phys.\ Lett.\ B {\bf 541}, 201 (2002)
  [arXiv:astro-ph/0204357].

\bibitem{seealso} 
M. Livio and M. Stiavelli, 
  Ap. J. Lett. {\bf 507} (1998) L13
  [arXiv:astro-ph/9808291];
S.~J.~Landau and H.~Vucetich,
  Astrophys.\ J.\  {\bf 570}, 463 (2002)
  [arXiv:astro-ph/0005316].

\bibitem{op1}
K.~A.~Olive and M.~Pospelov,
  Phys.\ Rev.\  D {\bf 65}, 085044 (2002)
  [arXiv:hep-ph/0110377].

\bibitem{cststring}
T.~Damour and A.~M.~Polyakov,
  Nucl.\ Phys.\  B {\bf 423}, 532 (1994)
  [arXiv:hep-th/9401069];
T.~Damour, F.~Piazza and G.~Veneziano,
  Phys.\ Rev.\  D {\bf 66}, 046007 (2002)
  [arXiv:hep-th/0205111];
T.~Damour and J.~F.~Donoghue,
  Phys.\ Rev.\  D {\bf 82}, 084033 (2010)
  [arXiv:1007.2792 [gr-qc]].

\bibitem{Petitjean}
H.~Chand,  {\it et al.},
  Astron.\ Astrophys.\  {\bf 417}, 853 (2004)
  [arXiv:astro-ph/0401094];
R.~Srianand,  {\it et al.},
  Phys.\ Rev.\ Lett.\  {\bf 92}, 121302 (2004)
  [arXiv:astro-ph/0402177].
  
\bibitem{quast}
R.~Quast, D.~Reimers and S.~A.~Levshakov,
  Astron.\ Astrophys.\  {\bf 415}, L7 (2004)
  [arXiv:astro-ph/0311280].

  \bibitem{sys}
T.~Ashenfelter, G.~J.~Mathews and K.~A.~Olive,
  Phys. Rev. Lett. {\bf 92}, 041102 (2004);
T.~P.~Ashenfelter, G.~J.~Mathews and K.~A.~Olive,
  Astrophys. J.  {\bf 615}, 82 (2004);
M.~G.~Kozlov, {\it et al.},
  Phys. Rev. A {\bf 70}, 062108 (2004).

  \bibitem{chameleon}
J.~Khoury and A.~Weltman,
 Phys.\ Rev.\ Lett.\  {\bf 93}, 171104 (2004)
 [arXiv:astro-ph/0309300];
 K.~Hinterbichler and J.~Khoury,
 Phys.\ Rev.\ Lett.\  {\bf 104}, 231301 (2010)
 [arXiv:1001.4525 [hep-th]].
  
  

\bibitem{ellisolive}
 J.~R.~Ellis,  {\it et al.},
  Phys.\ Lett.\  B {\bf 228}, 264 (1989).
  
\bibitem{op2}
K.~A.~Olive and M.~Pospelov,
  Phys.\ Rev.\  D {\bf 77}, 043524 (2008)
  [arXiv:0709.3825 [hep-ph]].
  
\bibitem{barrow}
B.~Li, D.~F.~Mota and J.~D.~Barrow,
  arXiv:1009.1396 [astro-ph.CO].

\bibitem{opu}
K.~A.~Olive, M.~Peloso and J.~-P.~Uzan,
  Phys.\ Rev.\ D {\bf 83}, 043509 (2011)
  [arXiv:1011.1504 [astro-ph.CO]].
  
  
\bibitem{wall-alpha} 
  T.~Chiba and M.~Yamaguchi,
  JCAP {\bf 1103}, 044 (2011)
  [arXiv:1102.0105 [astro-ph.CO]];
  K.~Bamba, S.~'i.~Nojiri and S.~D.~Odintsov,
  Phys.\ Rev.\ D {\bf 85}, 044012 (2012)
  [arXiv:1107.2538 [hep-th]].
 


\bibitem{gl}
J.~Gasser and H.~Leutwyler,
 Phys.\ Rept.\  {\bf 87} (1982) 77.


\bibitem{obswall}
 A. E. Everett, Phys. Rev. D {\bf 10} 3161 (1974);
 Ya. B. ZelÕdovich et al., Sov. Phys. JETP {\bf 40} 1 (1975). 
 
  
\bibitem{CMBstring}
  Y.~B.~Zeldovich, I.~Y.~Kobzarev and L.~B.~Okun,
  Zh.\ Eksp.\ Teor.\ Fiz.\  {\bf 67}, 3 (1974)
  [Sov.\ Phys.\ JETP {\bf 40}, 1 (1974)];
A.~A.~Fraisse, {\it et al.},
  Phys.\ Rev.\  D {\bf 78}, 043535 (2008)
  [arXiv:0708.1162 [astro-ph]];
  R.~Durrer, M.~Kunz and A.~Melchiorri,
  Phys.\ Rept.\  {\bf 364}, 1 (2002)
  [arXiv:astro-ph/0110348];
    G.~Rocha, {\it et al.} 
  Mon.\ Not.\ Roy.\ Astron.\ Soc.\  {\bf 352}, 20 (2004)
  [arXiv:astro-ph/0309211].
  M.~Nakashima, R.~Nagata and J.~Yokoyama,
  Prog.\ Theor.\ Phys.\  {\bf 120}, 1207 (2008)
  [arXiv:0810.1098 [astro-ph]];
 E.~Menegoni,  {\it et al.},
  Phys.\ Rev.\  D {\bf 80}, 087302 (2009)
  [arXiv:0909.3584 [astro-ph.CO]];
  C.~J.~A.~Martins,  {\it et al.},
  Phys.\ Rev.\  D {\bf 82}, 023532 (2010)
  [arXiv:1001.3418 [astro-ph.CO]].
  S.~J.~Landau and C.~G.~Scoccola,
  arXiv:1002.1603 [astro-ph.CO].

   \bibitem{walldyn}
 J.~C.~R.~Oliveira, C.~J.~A.~Martins and P.~P.~Avelino,
  Phys.\ Rev.\  D {\bf 71}, 083509 (2005)
  [arXiv:hep-ph/0410356].

  
\bibitem{Komatsu:2010fb}
  E.~Komatsu {\it et al.}  [WMAP Collaboration],
  Astrophys.\ J.\ Suppl.\  {\bf 192}, 18 (2011)
  [arXiv:1001.4538 [astro-ph.CO]].


%
\bibitem{Murphy:2003mi}
  M.~T.~Murphy, V.~V.~Flambaum, J.~K.~Webb, V.~V.~Dzuba, J.~X.~Prochaska and A.~M.~Wolfe,
  Lect.\ Notes Phys.\  {\bf 648}, 131 (2004)
  [arXiv:astro-ph/0310318].

%
\bibitem{Murphy:2009az} 
  M.~T.~Murphy, J.~K.~Webb and V.~V.~Flambaum,
  arXiv:0911.4512 [astro-ph.CO].

\bibitem{link}
http://adsabs.harvard.edu/abs/2009MmSAI..80..833M


\bibitem{Griest:2009rv} 
  K.~Griest, J.~B.~Whitmore, A.~M.~Wolfe, J.~X.~Prochaska, J.~C.~Howk and G.~W.~Marcy,
  Astrophys.\ J.\  {\bf 708}, 158 (2010)
  [arXiv:0904.4725 [astro-ph.CO]].



\bibitem{Kobzarev:1974cp} 
  I.~Y.~.Kobzarev, L.~B.~Okun and M.~B.~Voloshin,
  Sov.\ J.\ Nucl.\ Phys.\  {\bf 20}, 644 (1975)
  [Yad.\ Fiz.\  {\bf 20}, 1229 (1974)].


\bibitem{Coleman:1977py} 
  S.~R.~Coleman,
  Phys.\ Rev.\ D {\bf 15}, 2929 (1977)
  [Erratum-ibid.\ D {\bf 16}, 1248 (1977)].



  \bibitem{newflam}
   J.~C.~Berengut, E.~M.~Kava and V.~V.~Flambaum,
  arXiv:1203.5891 [astro-ph.CO].
  
  
\end{thebibliography}
\end{document}